\documentclass{emulateapj}
\usepackage{epsf}
\usepackage{apjfonts}
\usepackage{xspace}
\usepackage{amsmath}
\usepackage{framed} 

   \usepackage{graphicx}

\def\Cf{C_{{\rm HII}}}
\def\Cfloc{C^{\rm loc}_{\rm HII}}

\def\alphaA{\alpha_{{\rm A}}}

\def\fcoll{f_{\rm coll}}

\def\fesca{f_{\rm esc, rel, 0}}
\def\rfesc{f_{\rm esc, rel}}

\def\Mcrit{M_{\rm crit}}

\def\Msun{{\rm M}_\odot}

\def\ss{\sigma^2}

\def\HII{{\rm H\,II}}

\def\NH{N_{\rm H}}
\def\NgH{N_{\rm \gamma/H}}
\def\NiH{N_{\rm i/H}}

\def\dim#1{\mbox{\,#1}}

\begin{document}

\title{Recombination clumping factor during cosmic reionization}

\author{Alexander A.\ Kaurov\altaffilmark{1} and Nickolay Y.\ Gnedin\altaffilmark{2,1,3}}
\altaffiltext{1}{Department of Astronomy \& Astrophysics, The
  University of Chicago, Chicago, IL 60637 USA; kaurov@uchicago.edu}
\altaffiltext{2}{Particle Astrophysics Center, 
Fermi National Accelerator Laboratory, Batavia, IL 60510, USA; gnedin@fnal.gov}
\altaffiltext{3}{Kavli Institute for Cosmological Physics and Enrico
  Fermi Institute, The University of Chicago, Chicago, IL 60637 USA} 

\begin{abstract}
We discuss the role of recombinations in the IGM, and the related concept of the clumping factor, during cosmic reionization. The clumping factor is, in general, a local quantity that depends on the local over-density and the scale below which the baryon density field can be assumed smooth. That scale, called the filtering scale, is itself depended on over-density and local thermal history. We present a method for building a self-consistent analytical model of inhomogeneous reionization assuming the linear growth rate of the density fluctuation, which accounts for these effects simultaneously. We show that taking into account the local clumping factor introduces significant corrections to the total recombination rate, comparing to the model with a globally uniform clumping factor. 
\end{abstract}


\section{Introduction}
\label{sec:intro}

Studying cosmic reionization is first and foremost a counting exercise. Ionizing photons are being produced in sources, used up to ionize fresh hydrogen (in this paper we focus on hydrogen reionization, although helium reionization is qualitatively similar), and are being wasted in ionizing those of hydrogen atoms that managed to recombine after the first ionization. It is these, wasted photons, that are the subject of this study.

Let us consider some region of the universe, and let $\NiH$ be the number of ionizations per hydrogen nucleus required to keep the region ionized. If we ignore recombinations, then, obviously, $\NiH=1$. With recombinations
\begin{equation}
  \NiH(t) = 1 + \int_0^t \frac{dt}{\bar{t}_{\rm rec}},
  \label{eq:nih}
\end{equation}
where $\bar{t}_{\rm rec}$ is the average recombination time in the region.

If now $\NgH$ is the number of hydrogen ionizing photons per one hydrogen nucleus in this region that are either produced inside the region or arrive into it from the external sources, then the condition for the region to be reionized is simply
\begin{equation}
  \NgH=\NiH.
  \label{eq:reicond}
\end{equation}
Perhaps not surprisingly, this simple equation encapsulates most of existing models of reionization. For example, the original \citet{Furlanetto2004} model is a limiting case of Equation (\ref{eq:reicond}) with $\NiH={\rm const}=\langle\NiH\rangle$ and $\NgH$ is proportional to the fraction of all matter in the region that is collapsed into virialized objects times some photon production efficiency factor,
\begin{equation}
  \NgH/\langle\NiH\rangle \equiv \zeta m_{\rm coll}/m_{\rm tot} = 1.
  \label{eq:furl04}
\end{equation}
In this equation, the efficiency factor $\zeta$ can account for a whole variety of effects, including recombination in  Intergalactic Medium (IGM) and Lyman Limit systems (LLS). The disadvantage of this simple notation is that no matter what underlying physics is included in calculation of $\zeta$, it will still depend only on the collapsed fraction and not on the ionized fraction or bubble sizes, which is crucial for accurate treatment of inhomogeneous recombination and absorption by LLS.

Another commonly used model of reionization by \citet{igm:mhr99} is obtained from Equation (\ref{eq:reicond}) by differentiating with respect to cosmic time $t$ the both sides of the equation, averaged over the probability $Q_\HII$ that the region is ionized,
\[
  \frac{d}{dt}\langle\NgH Q_\HII\rangle = \frac{d\bar{Q}_\HII}{dt} + \frac{1}{\bar{t}_{\rm rec}}\bar{Q}_\HII.
\]

Yet another commonly used model of \citet[][see their Equation\ (5)]{igm:mhr00} is a generalization of Equation (\ref{eq:reicond}) with an additional term on the right hand side that accounts for ionizing photons stored in the cosmic background radiation and that expands the recombination time as an integral over the density PDF.

Given a source model, $\NgH$ can be computed. The challenge of modeling the spatially inhomogeneous reionization is then in accounting for recombinations in the second term of Equation (\ref{eq:nih}). In principle, one can model the average recombination time over any spatial region. In practice, this is virtually impossible because recombinations take place in greatly varied physical conditions.

Some of the ionizing photons that leave the surface of a massive star or the accretion disk around a black hole will be lost in the ISM of the parent galaxy. Following such recombinations is a nightmare, as it would require detailed ISM models of high redshift galaxies and quasars. Therefore such recombinations are commonly treated not as actual recombinations, but as imperfect escape of ionizing photons from the source, and quantified by the \emph{escape fraction} - i.e. these recombinations are moved to the left-hand-side of Equation (\ref{eq:reicond}). It is useful to remember, however, that the escape fraction actually quantifies (a portion of) photon loss due to recombinations.

After leaving the source, escaped ionizing photons move through space. Some of these photons will be expended on ionizing fresh hydrogen atoms (a unity factor in Equation \ref{eq:nih}); most of those that manage to fly unimpeded for a mean-free-path will be absorbed by LLS. Absorptions in LLS are also recombinations, but just like the photon loss in the source ISM, it is cumbersome to account for them by actually counting recombinations. Instead, it is more convenient to treat the effect of LLS as the finite mean-free-path for ionizing photons \citep{igm:m03,igm:sc10}. The importance of the absorption by minihalos (compact neutral objects) is discussed in \cite{Haiman2001}, where this effect is treated as an addition to the clumping factor of IGM. The distribution of LLS in semi-numerical and numerical simulation is studied in \citet{Crociani2011} and \citet{Alvarez2012}.

In the language of Equation (\ref{eq:reicond}) these recombinations are again counted as a reduction in $\NgH$ - photons from not all distant sources can reach a given place in the universe, but only from those located no more than a mean free path away. 

Yet, there will be some atoms in the general IGM that will recombine, and these extra recombinations should be accounted for by the second term of Equation (\ref{eq:nih}). Hence, Equation (\ref{eq:nih}) can be written as
\begin{equation}
  \NiH(t) = 1 + \frac{1}{\NH} \int_0^t dt \int_V \alpha(T) n_e n_\HII dV,
  \label{eq:nih2}
\end{equation}
where $\NH$ is the total number of hydrogen nuclei (ionized and neutral) in the volume $V$. Because the volume integral in Equation (\ref{eq:nih2}) cannot be evaluated directly without a highly detailed numerical simulation, it is often parametrized by a \emph{clumping factor} $\Cf$,
\begin{equation}
  \NiH(t) = 1 + \int_0^t dt \alphaA \Cf(t,V) \langle n_e\rangle_V \langle x_\HII\rangle_M,
  \label{eq:chii}
\end{equation}
where $\alphaA$ is some reference value for the recombination rate, and $\langle\rangle_{M,V}$ denote mass and volume average over the volume $V$ respectively. The definition of $\Cf$ then simply follows from the comparison of Equations (\ref{eq:nih2}) and (\ref{eq:chii}),
\[
  \Cf(t,V) \equiv \frac{\langle \alpha(T) n_e n_\HII\rangle_V}{\alphaA \langle n_e\rangle_V \langle n_\HII\rangle_V}.
\]

Hereafter, we choose the case-A recombination coefficient at $T=20{,}000\dim{K}$ as our reference recombination rate,
\[
   \alphaA = 2.51\times10^{-13}\dim{cm}^3/\dim{s}.
\]

In the Appendix (\S \ref{sec:coef}) we show that during reionization case A recombination photons are mostly redshifted out of resonance, and therefore the case A recombination coefficient is the proper choice for modeling reionization.

Changing $\alphaA$ rescales the clumping factor. Notice, however, that in the limit of the perfectly homogeneous universe the clumping factor $\Cf$ is not necessarily 1, since it depends on the actual recombination rate that may be different from the reference value. In particular, $\Cf$ can be less than 1 if $\alpha(T)$ is below $\alphaA$.

The paper is organized as follows. In the \S\ref{sec:analyt} we briefly describe the analytical model of reionization first introduced by \citet{Furlanetto2004} and the extension of this model by \citet{Kaurov2013}, which is important for this paper. In \S\ref{sec:clumping} the usual notation of the clumping factor is presented and compared with our approach. In \S\ref{sec:loc} we describe how the local clumping factor can be implemented. In \S\ref{sec:compare} we perform the comparison with previous works. We show the results in \S\ref{sec:results} and in the last section \S\ref{sec:concl} we conclude.

\section{Method}

\subsection{Analytical model of reionization}
\label{sec:analyt}

We use the analytical model of the epoch of reionization by \citet{Furlanetto2004}, which is based on the excursion set formalism. It provides information not only about the global ionization history, but also about the distribution of sizes of ionized bubbles. In \citet{Kaurov2013} this model was extended in order to match the actual observed galaxy luminosity functions, to account for galaxy biasing, and for absorption by LLS. We also included tracking of merging of separate ionized bubbles, which is necessary for the calculation of the recombination rate. In the language of Equation \ref{eq:furl04}, the modification we made in \citet{Kaurov2013} allow to make $\zeta$ the function of local merging and thermal history.

The main ingredient of any model of reionization is a model for the escape fraction of ionizing photons. We use a simple model with only two parameters -- the minimum mass $\Mcrit$ and the amplitude $\fesca$,
\begin{equation}
\rfesc(M) = \begin{cases}
\fesca, & M>\Mcrit \\
0, & M\leq \Mcrit.
\end{cases}
\label{eq:fesc}
\end{equation}
Note, that since the observed UV galaxy luminosity functions serve as the direct input of our model, we only need to know the relative escape fraction, i.e.\ the ratio of the escape fraction of ionizing photons to that of the UV photons. 

Throughout this paper we adopt the values $\fesca=0.2$ and $\Mcrit=10^9$. We use the source model based on abundance matching of the actual observed galaxy UV luminosity functions \citep{Bouwens2007,Bouwens2010} and theoretically computed halo mass functions \citep{Tinker2008}. Photon loss due to LLS is accounted for by extrapolating the observed abundance of LLS to higher redshifts. We use an analytical fit to lower redshift ($z<6$) observations of Ly$\alpha$ forest, provided by \citet{Songaila2010}:
\begin{equation}
\Delta l(z) = 50 \left[\dfrac{1+z}{4.5}\right] ^{-4.44}\;\rm{Mpc}.
\end{equation}
Since LLS affect only the late stages of reionization we assume that such an extrapolation is accurate enough. Hence, in contrast to the recombination clumping factor discussed in this paper, the photon loss due to LLS does not depend on the local thermal or reionization history. The absorption of photons by LLS in a given point at a particular redshift depends only on the size of bubble this point is located in. Accounting for bubble merging allows us to track the information about the moment of ionization at each point in the universe, and, therefore, implicitly calculate how many recombinations took place in each region. For more detailed description of source model and LLS implementation see \citet{Kaurov2013}.

The excursion set formalism is based on the idea of representing each point in the universe as a one-dimensional function (a random walk trajectory) of matter over-density versus the smoothing scale $R$ or, alternatively, the variance of the density field $\sigma_R^2$. The crucial ingredient in the excursion set formalism is a \emph{barrier}. The barrier is a one-dimensional function of the smoothing scale $R$ (or, again, the variance $\sigma^2_R$). The barrier is used to separate all trajectories into two classes: those that cross the barrier and those that do not. The two classes are then identified with some physical properties of the modeled objects. For example, in the Press-Schechter formalism \citep{PressSchetcher1974} all trajectories that cross the barrier are identified with collapsed objects, dark matter halos. In the \citet{Furlanetto2004} model trajectories that cross the barrier at each redshift are identified with spatial locations that have been already ionized.

\subsection{Clumping factor}
\label{sec:clumping}

The clumping factor is often used in analytical models of reionization for taking into account the inhomogeneities in the IGM \citep[c.f.][]{Madau1999, Kuhlen2012}.

 Let us take a look at the equation for the average recombination rate in a volume $V$:
\begin{equation}
\label{eq:rate}
  {\cal R} = \Cf \alphaA \left\langle n_{{\rm HII}}\right\rangle_V \left\langle n_{{\rm e}}\right\rangle_V,
\end{equation}
where $\left\langle n_{{\rm HII}}\right\rangle_V$ is the number density of hydrogen ions and $\left\langle n_{{\rm e}}\right\rangle $ is the number density of free electrons, which can be written in terms of ionized hydrogen density, assuming that the first ionized state of helium follows the ionization of hydrogen, 

\begin{equation}
\label{eq:ne}
\langle n_{{\rm e^{-}}}\rangle_V = \left(1+Y_{{\rm p}}/4X_{{\rm p}}\right)\langle n_{{\rm HII}}\rangle_V
\end{equation} 

Let us look closer at the origin of the clumping factor. In an approximation where we ignore temperature fluctuations and assume that helium is always ionized once together with hydrogen ($n_e \propto n_{\rm HII}$), the clumping factor $\Cf$ takes a more familiar form
\[
  \Cf = \frac{\langle n_{\rm HII}^2\rangle_V}{\langle n_{\rm HII} \rangle_V^2}.
\] 

In the excursion set formalism we consider the density field smoothed on different spatial scales $R$. The scale of smoothing is more conveniently quantified by $\sigma^2(R)$, the variance of the density field on scale $R$. In that framework the local clumping factor in a finite region of scale $R$ (rms density fluctuation $\sigma^2(R)$) with the mean overdensity $\bar\delta$ becomes the integral over the linear density PDF,
\begin{eqnarray}
\Cfloc(\bar\delta,\sigma^2)
 & = & \dfrac{1}{\sqrt{2\pi(\sigma_{\infty}^2-\sigma^2)}} \int_{-\infty}^{+\infty}\mathrm{d}\delta\, e^{-\dfrac{(\delta-\bar\delta)^2}{2(\sigma_{\infty}^2-\sigma^2)}}
\nonumber\\
& & \times\left(1+\delta_{\rm HII}\right)^2  ,
\label{eq:Clocsigma}
\end{eqnarray}
where $\sigma^2_{\infty}$ is the rms density fluctuation on the smallest scale (it must be finite, or the clumping factor becomes infinite), and $\delta_{\rm HII}(\delta)$ is the fluctuation in the ionized hydrogen density. One can interpret this integral as subdividing the region in smaller pieces of scale $\sigma^2_{\infty}$, each one of which is uniform, i.e.\ with a unit clumping factor. 

In the linear regime $\delta_{\rm HII} = \delta$, and integral in Equation \ref{eq:Clocsigma} can be taken analytically:
\begin{equation}
\label{eq:Cfloclin}
\Cfloc(\bar\delta,\sigma^2)
=(1+\bar\delta)^2+(\sigma_{\infty}^2-\sigma^2).
\end{equation} 

This is the lower limit since any model for remapping linear overdensities into non-linear on small scales \citep[e.g.][]{Carron2013} will predict higher clumping factor. Ideally one would use a full local Probability Density Function (PDF) $p(\delta|\bar\delta,...)$, which accounts for small scale structure, local overdensity and temperature history. However, calculating of such a PDF is equivalent to the high resolution simulation of reionization in a large cosmological volume (in order to cover regions with wide range of overdensities). This kind of simulation is not available at the moment \citep{Trac2011}.

Physical meanings of the two terms in Equation (\ref{eq:Cfloclin}) are rather obvious. The first term is the overall, large-scale change in the gas density inside an ionized bubble - in denser bubbles the recombination rate is enhanced (compared to the mean), in the under-dense bubbles recombinations it is suppressed. The second term accounts for the small-scale clumpiness of the gas inside the ionized bubble. In the linear approximation (\ref{eq:Cfloclin}) those two terms are independent and additive.

To extend the model even further, one may consider $\sigma_{\infty}^2$ as a function of overdensity $\bar\delta$. We discuss our particular choice for $\sigma_{\infty}^2$ in the next subsection.

Plugging the local clumping factor into the Equation (\ref{eq:rate}) the recombination rate in a particular ionized bubble ($Q_{{\rm HII}}=1$) becomes:
\begin{equation}
\label{eq:ratemod}
{\cal R}(\delta)= \alphaA\left\langle n_{{\rm H}}\right\rangle_V \left\langle n_{{\rm e^{-}}}\right\rangle_V
\left( (1+\delta)^2+(\sigma_{\infty}^2-\sigma^2) \right) 
\left(1-\fcoll\right)^{2},
\end{equation}
where $n_{{\rm e^{-}}}$ is taken from Equation (\ref{eq:ne}) and $\fcoll$ is the fraction of matter collapsed into galaxies and LLS (and, hence, not making the IGM). We add the term $\left(1-\fcoll\right)^2$ for completeness. It takes away all collapsed hydrogen atoms and electrons from $\left\langle n_{{\rm H}}\right\rangle_V$ and  $\left\langle n_{{\rm e^{-}}}\right\rangle_V$. In general, $\fcoll$ is a strong function of redshift, $\delta$ and $\ss$ (the scale of the region that we consider). In the simplest way it can be calculated by integrating the Press-Schechter mass function above some minimum mass for collapse \citep{Furlanetto2004}. This factor is negligible if we consider the universe on average; however, during the very early stages of reionization individual bubbles surrounding galaxies can be small, and, therefore, the collapsed fraction inside them might be significant.

Equation (\ref{eq:ratemod}), with the linear ansatz for the local clumping factor, is our fiducial model. Since in the linear regime the cross-correlation between the large-scale term $(1+\bar\delta)^2$ and the small-scale term $(\sigma_{\infty}^2-\sigma^2)$ is zero, our fiducial model always \emph{underestimates} the real clumping.

\subsection{Filtering scale}
\label{sub:fscale}

We associate the rms density fluctuation on the smallest scale $\sigma_{\infty}^2$ with the filtering scale $k_F$ over which the linear fluctuations in the baryonic component are suppressed by the gas pressure forces \citep{Gnedin1998a,Gnedin2003}. 

The filtering scale depends on the cosmological parameters and thermal history of the universe or a finite region inside the universe \citep{Gnedin1998a}. In order to define thermal history for each region, we assume that each point in space became ionized instantly. The temperature at that point is negligibly small until the moment of ionization, when it jumps to $2.5\times10^4 \rm K$ and cools down afterward as $(1+z)^{0.9}$ \citep{ng:hg97}.

\begin{figure}
\includegraphics{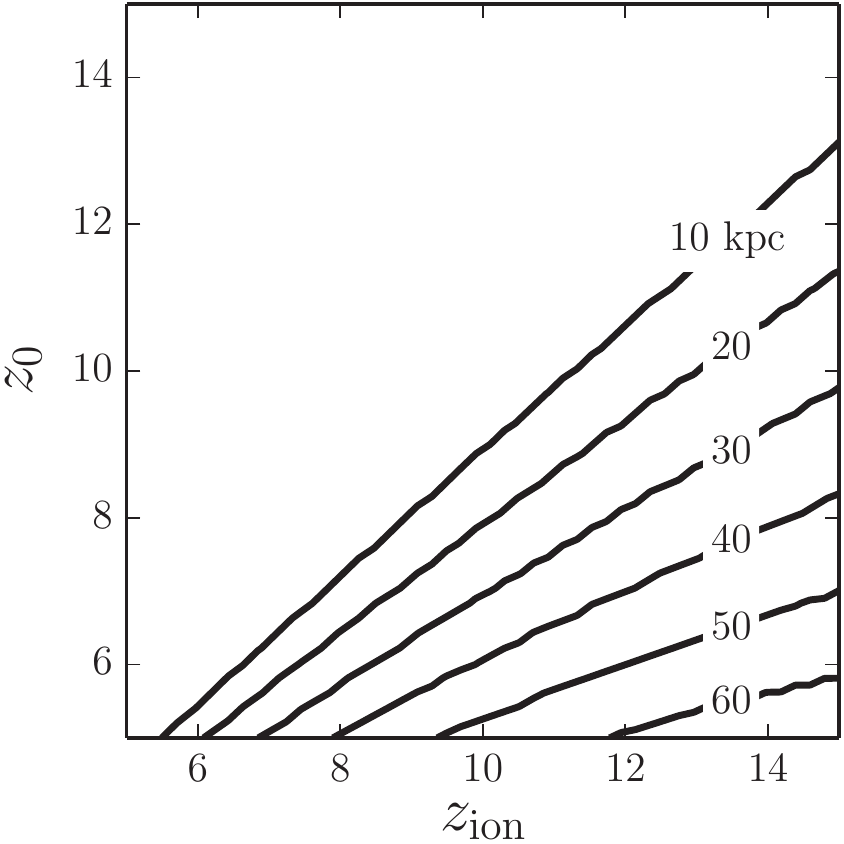}
\caption{The local filter scale at redshift $z_0$ of the region which was ionized at $z_{ion}$, assuming that the temperature is negligible before $z_{ion}$ and cools down afterward as $(1+z)^{0.9}$. }\label{fig:figap}
\end{figure}

In Figure \ref{fig:figap} the filtering scale is presented as a function of current redshift $z_0$ and ionization redshift $z_\mathrm{ion}$. 

\subsection{Applying local clumping factor}
\label{sec:loc}

Given the model for the thermal history, we compute $\sigma_{\infty}^2$ for each ionized bubble. The excursion set formalism allows us to build merger trees for ionized bubbles \citep{Furlanetto2005}. Integrating along the merger tree, we find volume averaged distribution of ionization redshifts inside bubbles of a specific scale.

In practice we find the barriers by performing the integration in redshift space. Beside the barrier itself, we keep two other pieces of information for each scale: the number of photons that were lost by recombination and absorption by LLS, and the distribution of the first time crossing redshifts, i.e. the ionization histories. At the next time step we numerically find a new barrier. Here we outline the basics of the algorithm:
\begin{enumerate}
  \item For a given $\sigma^2$ we make guess for $\delta$.
  \item Calculate the total number of produced photons inside the region of given scale $\sigma^2$ and overdensity $\delta$ using halo mass function and luminosity function. Reduce the number of photons due to escape fraction\footnote{This step is described in details in \citet{Kaurov2013}.}.
  \item Calculate the merger tree.
  \item Integrate the distribution of progenitors to find out how many photons were already lost due to LLS and recombinations, and also we find the distribution of ionization histories.
  \item Calculate the number of photons that are lost in the current time step. The losses due to LLS can be found using known mean free path. The distribution of ionization histories is used for calculating the $\sigma^2_{\infty}$ and therefore the total recombination rate.
  \item Finally, we have the number of produced photons and the number of lost photons due to LLS and recombinations. If the number of photons equal to the number of baryons plus calculated losses within given threshold, than we save this guess for $\delta$ as well the number of total lost photons and distribution of ionization history. If the this criteria is not met, we start from the beginning. 
\end{enumerate}

We do not discuss how we are making the guess for $\delta$ at step 1. It can be done in many intelligent ways in order to reduce number of iterations. The integration is started at redshift 15, assuming nothing was ionized at that moment. We found the convergence for time steps below 0.01 in redshift space and binning below 0.1 in $\sigma^2$ space. Monte Carlo simulations of random walks are used to calculate merger trees and total ionized fractions.

\subsection{Comparison with previous works}
\label{sec:compare}

Photon loss due to recombinations has been first implemented in the barrier-based family of analytical models of inhomogeneous reionization in \citet{Furlanetto2005}. Here we highlight the differences with our approach.

\citet{Furlanetto2005} treat both processes, recombinations in the IGM and photon loss due to mini-halos, as a limit on the mean free path (MFP) of ionizing photons. They incorporate it in their model by introducing a second barrier, which accounts for a characteristic scale at which bubbles stop growing. These approach leads to the abrupt cut-off of bubble size distribution, and ignores a possibility that two neighboring bubbles, each with size less than MFP, can merge together to a larger-than-MFP bubble.

In \citet{Kaurov2013} we presented our approach of implementing the photon loss due to LLS, which does not require a second barrier and does not introduce abrupt features in the bubble size distribution. In this paper we account for the recombinations in the IGM through a local (i.e.\ spatially variable and evolving) clumping factor; the effect of IGM clumping on the bubble size distribution is then modeled self-consistently as a balance between the production of ionizing photons and their consumption in ionizing both fresh neutral gas and previously recombined ionized gas.

In another class of analytical models \citep{Robertson2010,Kuhlen2012} the clumping factor is treated as a global external parameter, which needs to be deduced from other studies. Such an approach introduces three separate problems. First, the value of clumping factor is either chosen arbitrary, or adopted from numerical studies that may or may not be satisfying existing observational constraints. Second, the clumping factor is assumed to be uniform across the whole universe, taking the same value inside each ionized bubble. Third, such an approach is not self-consistent in a sense that there is no feedback from the progress of reionization on the value of the clumping factor. In this work we apply local clumping in analytical calculations of inhomogeneous reionization for the first time, and we explicitly calculate the clumping factor inside a region as a function of local reionization history. The local clumping factor, in turn, alters the local recombination rate, and, consequently, affects the local reionization history. This makes our model of reionization internally consistent with the value of the clumping factor used.

\begin{figure}
\includegraphics[width=1\columnwidth]{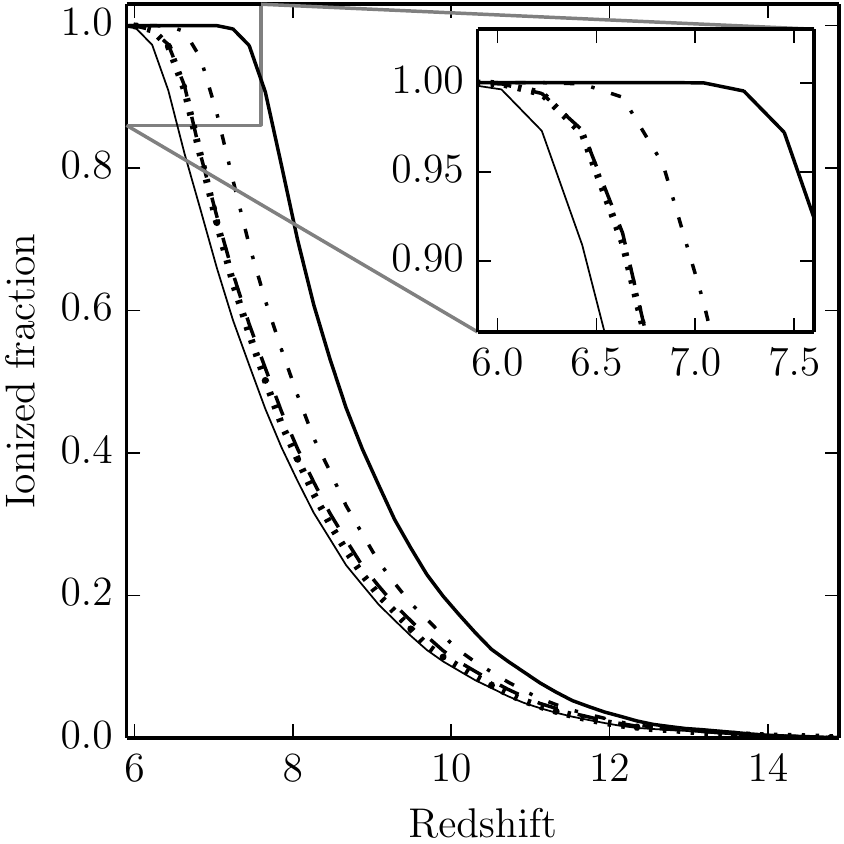}
\caption{Ionized fraction of hydrogen as a function of redshift for different recombination models. Thick solid line corresponds to the model without recombination in IGM (only LLS). Dot-dashed line calculated  with the clumping factor equal unity everywhere. Dashed line corresponds to the model without the large-scale contribution to the clumping factor in Equation (\ref{eq:ratemod}), $\bar\delta=0$. Dotted line is for the same model, but with the small-scale term $(\sigma_\infty^2-\sigma^2)$ set to zero. Thin solid line corresponds to the full model with both terms included. All models are calculated with fixed parameters: $\Mcrit=10^{9}\,\Msun$ and $\fesca=0.2$.
}
\label{fig:ionfrac}
\end{figure}

\section{Results}
\label{sec:results}

First we explore the effect of the local clumping factor on the global reionization history. Five models that are displayed in Figure \ref{fig:ionfrac} include: a model without any recombinations, a model without the clumping factor ($\Cf=1$), two models with each of the large- and small-scale terms removed in Equation (\ref{eq:Cfloclin}), and, finally, our fiducial model with full accounting for $\Cfloc$. 

\begin{figure}
\includegraphics[width=1\columnwidth]{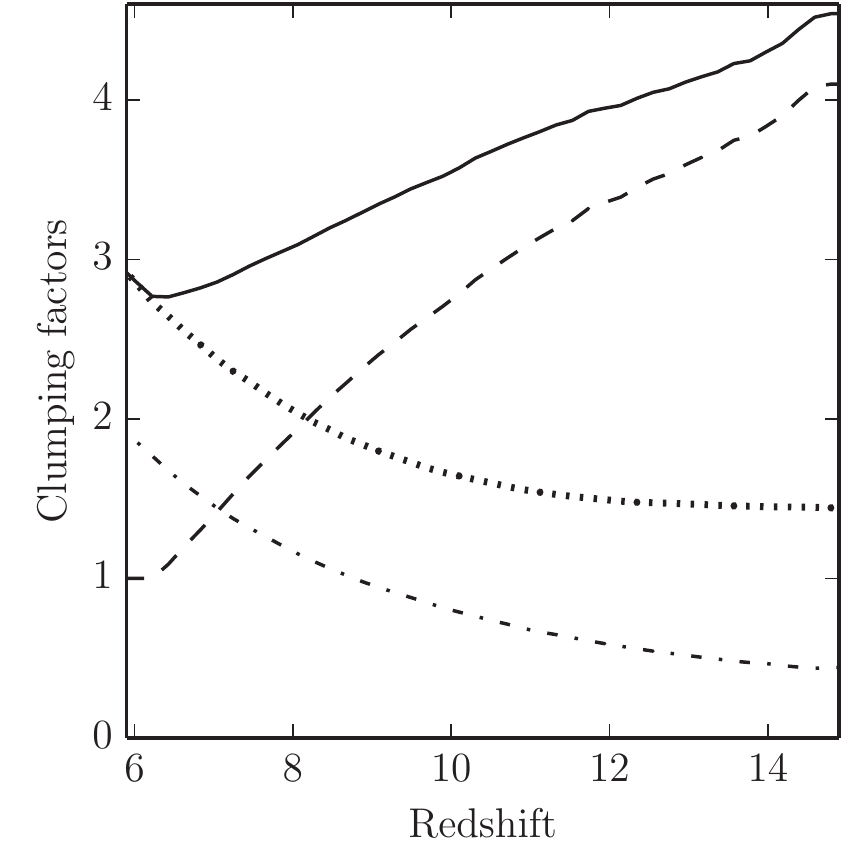}
\caption{Average clumping factor of ionized regions versus redshift for the model shown in Figure\ \ref{fig:ionfrac} with the solid thick line. Solid line shows the total value of the clumping factor from Equation (\ref{eq:Cfloclin}). The dashed line shows the contribution of the large-scale term $(1+\bar\delta)^2$ term averaged over all ionized regions. The dot-dashed line tracks the small-scale contribution $(\sigma_{\infty}^2-\sigma^2)$. The dotted line represents the total (not only in ionized regions) gas clumping factor $C_{b}=1+\langle\sigma_{\infty}^2\rangle$.}
\label{fig:Clumping}
\end{figure}

By coincidence the effects on the global reionization history from the overall change in the mean density of the ionized bubble and from the gas clustering inside the bubble are approximately the same (dotted and dashed lines in Figure \ref{fig:ionfrac}). However, their contributions to the clumping factor are different. They are further explored in Figure \ref{fig:Clumping}, where we show the evolution of the averaged over the whole universe clumping factor, as well as evolution of the spatially averaged large- and small-scale contributions. Notice that the averaging is done over the ionized regions only, so the spatial average of the $(1+\bar\delta)^2$ term from Equation (\ref{eq:Cfloclin}) is not identical to the one plus the rms value of the gas density.

Since at the end of reionization the mean densities of ionized bubbles approach the cosmic mean, with time the contribution of the large-scale, $(1+\bar\delta)^2$ term decreases, while the growth of structure causes the small-scale, $\sigma_\infty^2-\sigma^2$ term to increase.

The average over the whole universe clumping factor follows the clumping factor of ionized regions at smaller redshifts, since the universe is already mostly ionized. At higher redshifts it is, however, substantially lower, since higher density (i.e.\ more biased) regions reionize first. For example, at $z\sim15$ only about $3-\sigma$ regions are ionized, resulting in the recombination clumping factor being a factor of several higher than the total gas clumping factor.

The value of clumping factor at $z=6$ in recent numerical studies \citep{Pawlik2009, McQuinn2011, Shull2012, Finlator2012} varies between $2$ and $4$ with average around $3$ (see Figure (1) in \citet{Finlator2012} for comparison). The value we obtain is $C_b = 2.9$ (Figure \ref{fig:Clumping}). This comparison is not completely fair, however. In the simulations the clumping factor is usually computed in the regions with density below a given density threshold, which serve as an approximation to the general IGM. In our model recombinations in the IGM are counted by explicitly excluding recombinations local to a source (imposing the escape fraction) and recombinations in the LLS. Whether the two definitions of the general IGM are sufficiently compatible has not yet been tested properly.

\begin{figure}
\includegraphics[width=1\columnwidth]{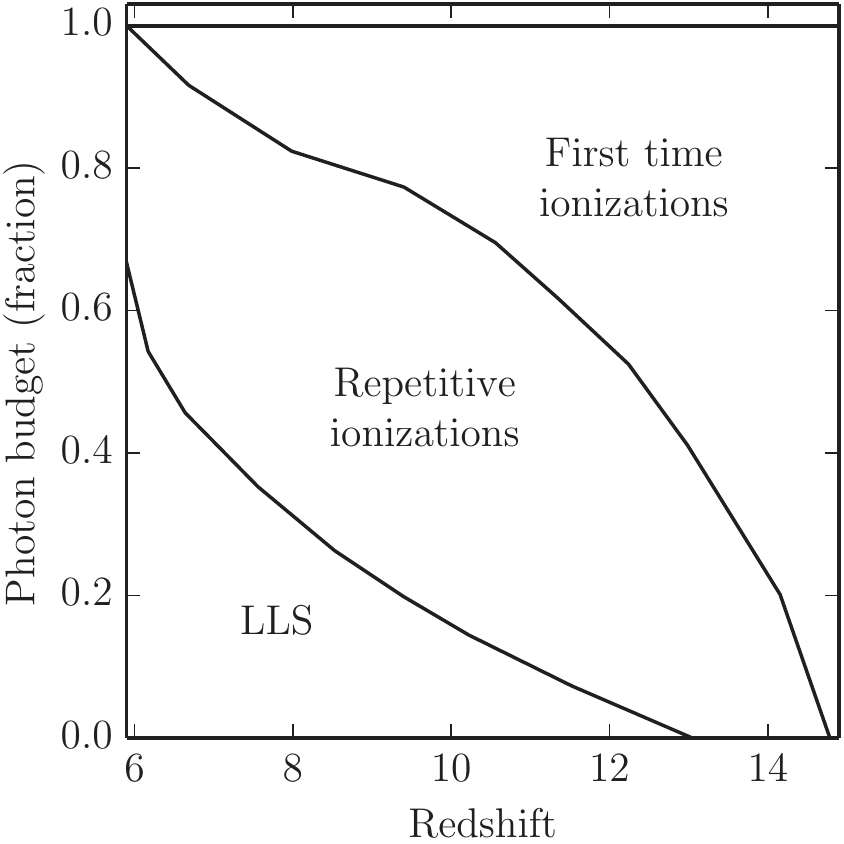}
\caption{Consumption of emitted photons by three processes as a function of redshift. Those are the first time ionizations, the secondary ionizations, which compensate recombination in IGM, and absorption by LLS.}
\label{fig:budget}
\end{figure}

Since in our analytical model the absorption by LLS and recombinations in IGM are well separated, we can plot the photon budget as a function of redshift. Figure \ref{fig:budget} shows the fractions of emitted photons, which were absorbed in different processes at each redshift. \textit{First time ionizations} of neutral atoms at the ionization fronts are shown on top. Their fraction decreases with time, because other processes become relatively more important. The fraction of photons spent on maintaining ionized regions is labeled as \textit{repetitive ionizations}, and their fraction increases as the ionized volume grows. The third contribution is the photon loss i \textit{LLS}, which dominates at late stages and after the reionization, but is negligible when bubbles are small.

\begin{figure}
\includegraphics[width=1\columnwidth]{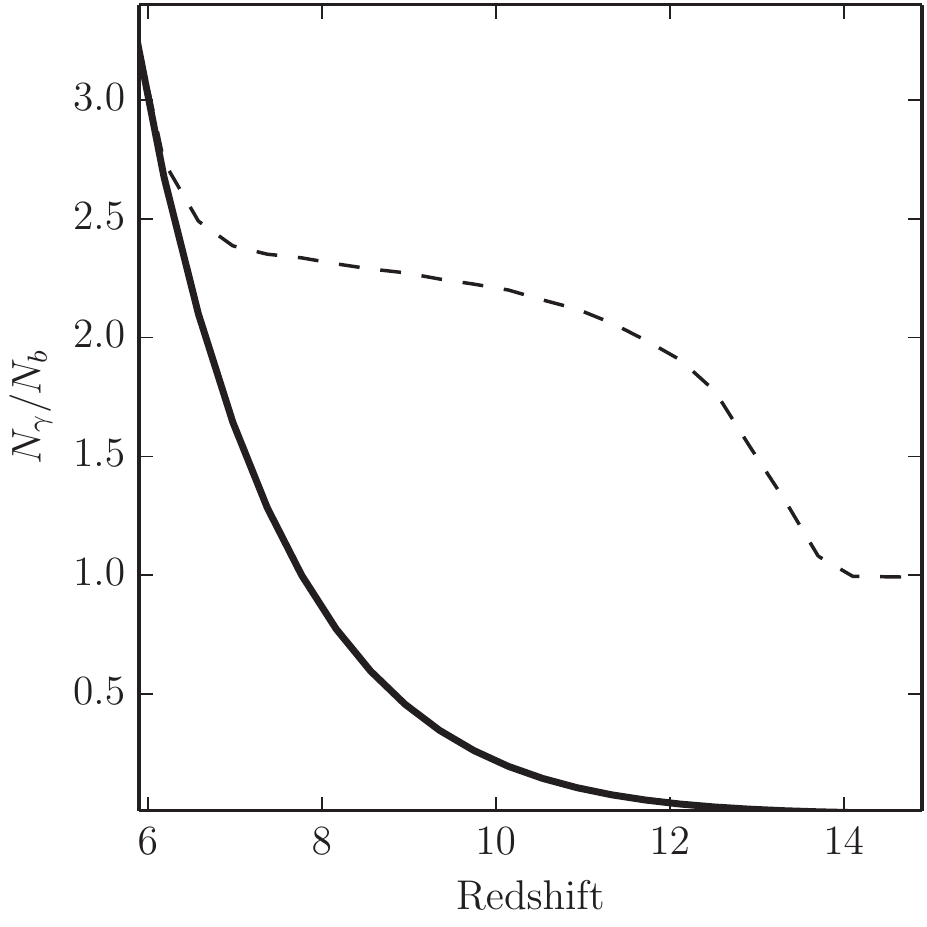}
\caption{\textbf{Number of ionizing photons escaped from all halos per baryon (solid line) and per ionized hydrogen (dashed line).}}
\label{fig:NgNb}
\end{figure}

Additionally, in Fig. \ref{fig:NgNb}, we show the total number of photons escaped from galaxies per baryon and per ionized baryon as a function of redshift. The deviation of the dashed line from unity represents the number of additional photons, that are required to overcome the combined effect of recombinations in the IGM and LLS. Figures \ref{fig:budget} and \ref{fig:NgNb} reveal that the ``effectiveness of reionization'' (i.e.\ the probability of a given ionizing photon to ionize a fresh hydrogen atom) is steadily declining with time.

\section{Conclusions}
\label{sec:concl}

The ionization photon losses during cosmic reionization are associated with spontaneous recombination. Recombinations can be subdivided into three categories. The first is recombinations inside galaxies and their hosting halos; such, internal to the source photon losses are commonly parametrized by the escape fraction.

The other two categories are recombinations in the general IGM and in the Lyman Limit systems. Recombinations inside LLS are more intuitively described by the finite mean free path of ionizing radiation, rather than by explicit count of individual recombinations.

The recombination rate in the IGM is proportional to the second power of density, and therefore is highly sensitive to the local overdensity and clumpiness of gas. In this paper we complement the \citet{Furlanetto2004} semi-analytic model of cosmic reionization, as extended by \citet{Kaurov2013}, with the treatment of the time- and spatially-dependent local clumping factor, that allows us to model the rate of recombination in the general IGM (in addition to already accounted for escape fraction and LLS).

The local clumping factor inside an ionized bubble can be split into two separate contributions: the "large-scale" one due to the overall variation in the mean density inside an ionized region, and a "small-scale" contribution due to gas clumpiness inside the ionized bubble. Both contributions can be correlated, enhancing the total clumping of the gas; they both depend on the local ionization and thermal histories, which control the evolution of the filtering scale and the small-scale distribution of inhomogeneities in the gas.

While for reionization histories consistent with the existing observational constraints the numeric value of the average clumping factor is not large (a few), it postpones the end of reionization (as compared to models without accounting for gas clumping) by a non-trivial amount. It is important to re-iterate that the recombination clumping factor is computed over the distribution of ionized bubbles \emph{only}, and as such is substantially larger than the total baryonic clumping factor $C_{b}=1+\langle\sigma_{\infty}^2\rangle$ (the latter does not enter the ionization balance equation and, therefore, does not directly affect the progress of reionization). Since higher density regions are more biased, and, hence, contain disproportionally larger fraction of ionizing sources, they reionize first, leading to higher clumping of ionized gas (as compared to the total gas) at earlier times.

\acknowledgements 
Fermilab is operated by Fermi Research Alliance, LLC, under Contract
No.~DE-AC02-07CH11359 with the United States Department of Energy.
This work was also supported in part by the NSF grant AST-1211190 and by the NASA grant NNX-09AJ54G. This work made extensive use of the NASA Astrophysics Data System and {\tt arXiv.org} preprint server. This work was done with significant usage of CosmoloPy Python package\footnote{http://roban.github.com/CosmoloPy/}.

\appendix
\section{Recombination coefficient}
\label{sec:coef}

The hydrogen recombination to the ground level produces a photon which can ionize another atom or can be absorbed by different processes. Two limiting cases are usually considered. Case A corresponds to the optically thin environment, where all recombination photons escape from the system. In the opposite, optically thick case B, all these photons are immediately absorbed, ionizing nearby hydrogen atoms.

During reionization four things can happen to a recombination photon:
\begin{enumerate}
\item it can be redshifted out of resonance, ceasing to be an ionizing photon;
\item another neutral hydrogen atom inside the ionized bubble can absorb it;
\item it can reach the boundary of the ionized bubble and ionize an atom in the still neutral IGM, thus contributing to the process of reionization; and
\item it can be absorbed by a LLS.
\end{enumerate}

In this section we compare the characteristic spatial scales of all these processes.

\begin{figure*}
\includegraphics[]{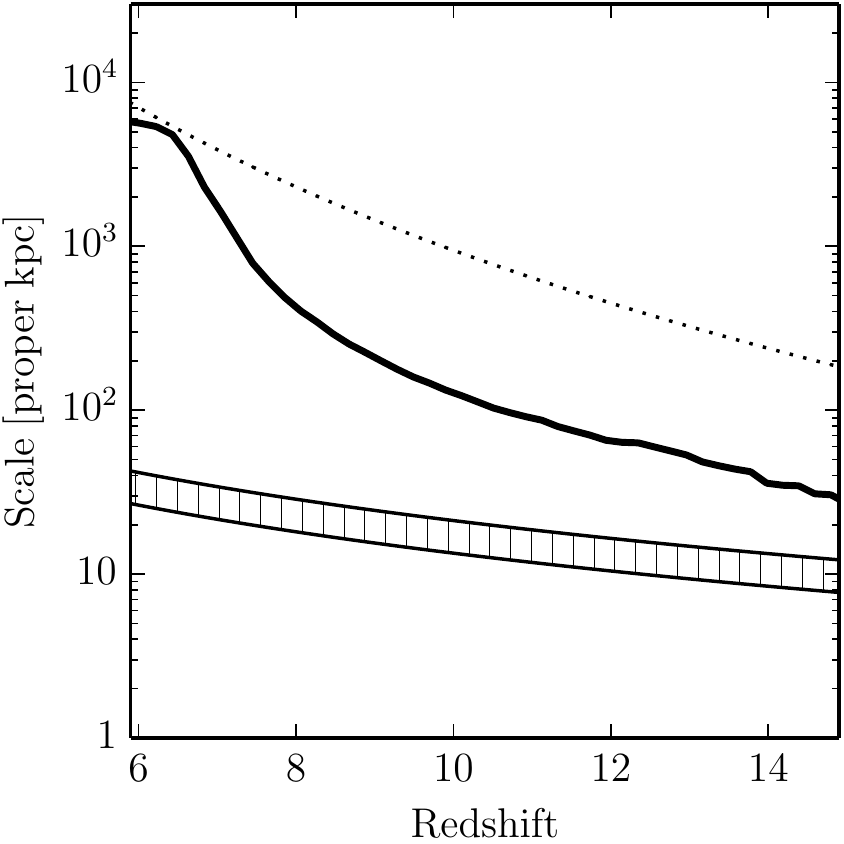}%
\;\;\;\;\;
\includegraphics[]{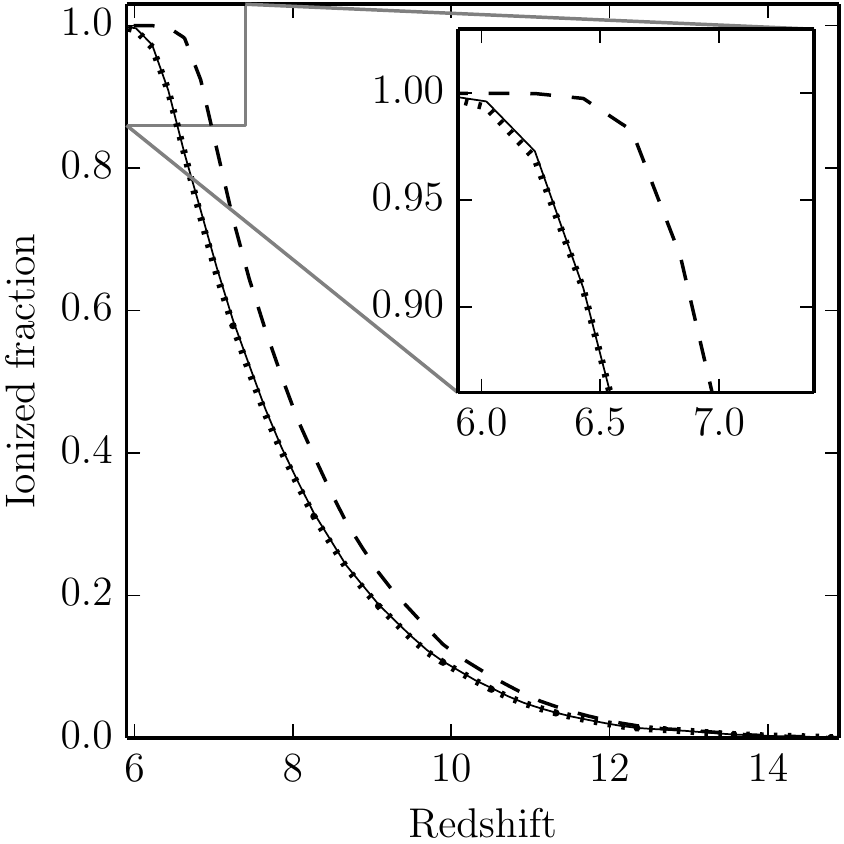}
\caption{\emph{Left}: Comparison of several spatial scales. The out of resonance scale for temperatures in range $(1..2.5)\times10^4\,K$ is presented by region with vertical hatching. Solid line represents the mean size of ionized bubbles in our fiducial model. Dotted line corresponds to the mean free path due to LLS absorption from \citet{igm:sc10}. \emph{Right}: Ionized fraction of hydrogen as a function of redshift for three models with different recombination coefficients. Dotted line corresponds to $\alpha_{\rm B}$, dashed line -- $\alpha_{\rm A}$, solid line -- $\alpha_{\rm X}$ (see Equation \ref{eq:alphaX}).}
\label{fig:scalecomp}
\end{figure*}

In order to estimate the light travel distance before getting out of resonance, we assume the typical temperature of the environment to be $T$. The line broadening due to Doppler effect can be estimated as:
\begin{equation}
\sigma_\nu=\sqrt{(kT)/(m_{{\rm p}}c^2)} \nu_0.
\end{equation}

To get out of this line, the photon should travel distance $D_\mathrm{R}$:

\begin{equation}
D_\mathrm{R} = ​\dfrac{2 \sigma_\nu}{H(z) \nu_0},
\end{equation}
factor of 2 appears because both the emitting and absorbing atoms have temperature $T$, $H(z)$ is the Hubble constant at redshift $z$.

The second effect, absorption inside an ionized bubble, can not be easily estimated, because a photon crosses a large region with variable ionization fraction and density. The framework of our analytical model does not provide any information about local ionization fraction (it is assumed to be 0 or 1). However, for as long as the ionized bubble continues to grow, inside-the-bubble absorption should be small,  or the bubble would not be growing. Hence, we neglect this process in our model.

The scale of the third process is set by the typical bubble size, which is directly predicted by our model. We plot this scale in Figures \ref{fig:scalecomp} with the solid line. Finally, the mean free path due to LLS absorption can be extrapolated to the redshifts of interest from the fits given in \cite{igm:sc10}.

In Figure \ref{fig:scalecomp} we compare these scales. The temperature $T$ of ionized medium is varied in the range $1-2.5 \times 10^4 K$. We can immediately conclude that LLS do not play any role in absorption of recombination photons until the end of reionization, because the mean free path due to them is the longest characteristic scale.

The mean bubble size is larger than the light travel distance before redshifted out of resonance at all redshifts, and, consequently, the mean free path in the ionized IGM is also longer. Therefore we may conclude that all recombination photons are redshifted out of resonance before they have a chance to ionize another atom. 

However, the average bubble size says nothing about the actual size distribution. Some bubbles might still be small enough for photons not being able to redshift enough before they hit the bubble wall. We can estimate this effect by comparing three different variations of our fiducial model. Two of these variations are with uniform recombination coefficients $\alpha_{\rm A}=2.51\times10^{-13}\,\rm cm^2$ and $\alpha_{\rm B}=1.43\times10^{-13}\,\rm cm^2$ for case A and B recombination. The third variation accounts for the bubble size in a following manner:
\begin{equation}
\label{eq:alphaX}
\alpha_{\rm X} = \alpha_{\rm B} + F(D_{\rm R}/D_{\rm bubble})\times(\alpha_{\rm A}-\alpha_{\rm B}),
\end{equation}
where $D_{\rm bubble}$ and $D_{\rm R}$ are the characteristic size of a bubble and the light travel distance before it is redshifted out of resonance correspondingly. The function $F(x)$ can be simply $\exp(-x)$ or a more complicated function that accounts for the distribution of recombination inside the bubble (see Appendix in \citet{Kaurov2013}). We found that for any reasonable choice of $F(x)$ the fraction of photons that ionize another atoms before they redshifted out of resonance is negligible. The result is shown in the right panel of Figure \ref{fig:scalecomp}, where solid line, which corresponds to the model with recombination coefficient from Equation \ref{eq:alphaX}, lies almost exactly on top of dashed line, the model with constant coefficient $\alpha_{\rm A}$.


\bibliographystyle{apj}
\bibliography{clump-hudf,refs,ng-bibs/igm,ng-bibs/self,ng-bibs/dsh,ng-bibs/rei}


\end{document}